\documentclass[11pt,a4paper]{article}
\usepackage[T1]{fontenc}
\usepackage[latin9]{inputenc}
\usepackage{amsmath}
\usepackage{graphicx}

\makeatletter

\pdfoutput=1 

\usepackage{jcappub}

\usepackage{bm}
\usepackage{cancel}
\def\bea{\begin{eqnarray}}
\def\eea{\end{eqnarray}}

\title{\boldmath Lensing contribution to the 21cm intensity bispectrum}

\author[a]{Rahul Kothari,}
\author[a,b]{Roy Maartens}

\affiliation[a]{Department of Physics and Astronomy, University of the Western Cape, South Africa}
\affiliation[b]{Institute of Cosmology \& Gravitation, University of Portsmouth, Portsmouth PO1 3FX, UK}

\emailAdd{quantummechanicskothari@gmail.com}
\emailAdd{roy.maartens@gmail.com}

\abstract{Intensity maps of the 21cm emission line of neutral hydrogen are lensed by intervening large-scale structure, similar to the lensing of the cosmic microwave background temperature map.
We extend previous work by calculating the lensing contribution to the full-sky 21cm bispectrum in redshift space. The lensing contribution tends to peak when  equal-redshift fluctuations are lensed by a lower redshift fluctuation. {At high redshift},  lensing effects can become comparable to the contributions from density and redshift-space distortions.}
\makeatletter
\gdef\@fpheader{}
\makeatother
\notoc
\begin{document}
\maketitle \flushbottom

\section{Introduction\label{intro}}
The cosmic microwave background (CMB) has been {an} invaluable {probe} for developing and testing cosmological models. Its main constraining power {comes} from the primary anisotropies that are imprinted at $z\sim 1000$. {In addition to this,}  it also contributes to low-redshift constraints via the lensing of the CMB temperature by large-scale structure  \cite{Aghanim:2018oex}.
The integrated 21cm emission from neutral hydrogen (HI) in the post-reionisation era produces maps that are qualitatively similar to the CMB, but with multiple maps over a range of redshifts.   21cm intensity maps are also lensed by intervening large-scale structure. For surveys that detect individual galaxies, the lensing effect on number density occurs at first order in perturbations and modifies the tree-level power spectrum. In the case of the CMB and  21cm intensity mapping, the first-order lensing effect vanishes due to conservation of surface brightness \cite{Hall:2012wd,Alonso:2015uua}: the lensing effect in the CMB and 21cm intensity arises at second order. As a result, the 21cm power spectrum is only affected at 1-loop level \cite{Umeh:2015gza,Jalivand:2018vfz}. By contrast, the tree-level 21cm bispectrum does carry an imprint of lensing, as pointed out in \cite{DiDio:2015bua,Jalivand:2018vfz,Durrer:2020orn}. 

In this paper, our aim is to derive the lensing contribution to the 21cm angular bispectrum and present {some} numerical examples. The result includes redshift-space distortions (RSD). Since we work in angular harmonic space, wide-angle correlations are naturally included, i.e., there is no flat-sky approximation.

The HI temperature contrast observed in redshift space is denoted $ \Delta\equiv\Delta_{\mathrm{HI}}=\Delta T_{\mathrm{HI}}/\big\langle T_{\mathrm{HI}} \big\rangle$. 
The lensed temperature contrast at redshift $z$ and in direction $\bm{n}$ is related to the unlensed one as
\begin{equation}
    \Delta^{\rm L}(z,\bm{n})=\Delta\big(z,\bm{n}+\bm\nabla_{\!\perp} \phi(z,\bm n)\big),
    \label{eq:LenHIContra}
\end{equation}
where   $\bm{\nabla}_{\!\perp}$ is the gradient operator on the 2-sphere orthogonal to $\bm{n}$, and $\phi$ is the lensing potential. At first order (which is all that  is needed for the tree-level bispectrum),
\begin{equation}
    \phi^{(1)} = -\int_{0}^{r}\mathrm{d}\tilde{r}\, \frac{\big(r-\tilde{r}\big)}{r\tilde{r}}\Big[\Phi^{(1)}+\Psi^{(1)}\Big],
    \label{lp}
\end{equation}
 where $r$ is the comoving line-of-sight distance and the metric potentials  in Poisson gauge  are given by {(neglecting vector and tensor modes)}
\begin{equation}
    \mathrm{d}s^2=a^2\big[-(1+2\Psi)\mathrm{d}\eta^2+(1-2\Phi)\mathrm{d}\bm{x}^2\big].
\end{equation}

At first order, \eqref{eq:LenHIContra} implies that $\Delta^{\rm L(1)}=\Delta^{(1)}$, so that up
to second order we have 
\begin{equation}\label{del2l}
\Delta^{\rm L}(z,\bm{n})=\Delta^{(1)}(z,\bm{n})+\Delta^{(2)}(z,\bm{n})-\big\langle\Delta^{(2)}\big\rangle(z)+L^{(2)}(z,\bm{n})-\big\langle L^{(2)}\big\rangle(z)\,,
\end{equation}
where our convention is $\Delta=\Delta^{(1)}+\Delta^{(2)}$ and
we have subtracted averages in order to ensure that $\big\langle\Delta^{\rm L}(z,\bm{n})\big\rangle=0$. The unlensed temperature contrasts  are \cite{Umeh:2015gza,DiDio:2015bua,Jalivand:2018vfz}
\bea
\Delta^{(1)} &=&b_{1}\delta^{(1)}+{1\over\mathcal{H}} \partial_{r}^{2}V^{(1)}\,,
\label{eq:DeltaFirOrd}\\
\Delta^{(2)} & =&b_{1}\delta^{(2)}+\frac{1}{2}b_{2}\big[\delta^{(1)}\big]^{2}+b_{s}s^{2}\notag\\
&&{} +{1\over\mathcal{H}}\partial_{r}^{2}V^{(2)}
+ {1\over\mathcal{H}^2}\Big(\big[\partial_{r}^{2}V^{(1)}\big]^{2}+\partial_{r}V^{(1)}\partial_{r}^{3}V^{(1)}\Big)
+{1\over\mathcal{H}}\Big[\partial_{r}V^{(1)}\partial_{r} \delta^{(1)}+{\delta^{(1)}\partial_{r}^{2}V^{(1)} }\Big]\!,~~~~
\label{eq:DeltaSecOrd}
\eea
where $\delta$ is the matter density contrast, $\partial_r=\bm n\cdot\bm\nabla$ and the velocity potential is defined so that the peculiar velocity is $\bm v=\bm \nabla V$. Terms with radial gradients of $V$ constitute the RSD contribution.
The linear and quadratic clustering bias parameters are assumed scale-independent, i.e., $b_i=b_i(z)$. The tidal contribution to clustering bias has bias parameter $b_s(z)$ {multiplying}   
$s^2=s_{ij}s^{ij}$, where the tidal field is 
\begin{equation}
s_{i j}=\left(\partial_{i} \partial_{j}-\frac{1}{3} \delta_{i j} \nabla^{2}\right) \nabla^{-2} \delta^{(1)}\,.
\end{equation}

In the case of galaxy surveys,
the lensing contribution to number count fluctuations at first and second orders includes the lensing convergence,
\begin{equation}\label{lcon}
    \kappa=-\frac{1}{2}\nabla_{\!\perp a}\nabla^{a}_{\!\perp} \phi\,.
\end{equation}
By contrast, lensing of HI intensity fluctuations at  leading order (i.e. second order) does {\em not} include the lensing convergence. Instead, it is given  purely by a coupling of the lensing deflection angle $\nabla_{\!\perp}^{a}\phi$
with the screen-space gradient of the observed temperature contrast $\nabla_{\!\perp a}\Delta$ \cite{Umeh:2015gza,DiDio:2015bua}:
\begin{equation}
L^{(2)}(z,\bm{n})=\nabla_{\!\perp}^{a}\phi^{(1)}(z,\bm{n})\,\nabla_{\!\perp a}\Delta^{(1)}(z,\bm{n})\,.\label{eq:Lens_Contri}
\end{equation}

The same form of lensing contribution arises in the CMB. However, in the CMB case, the coupling in \eqref{eq:Lens_Contri} is {negligible, since there is effectively no correlation between primary CMB temperature fluctuations $\nabla_{\!\perp a}\Delta^{(1)}_{\rm cmb}$,  that are generated
at $z\sim1000$, and  the lensing deflections $\nabla_{\!\perp}^{a}\phi^{(1)}$, that are induced by the large-scale structure at low $z$   \cite{Jalivand:2018vfz} (see the review \cite{Lewis:2006fu} for further details). This correlation is not negligible
for post-reionisation 21cm intensity mapping, since the fluctuations $\nabla_{\!\perp a}\Delta^{(1)}$ are growing after reionisation, i.e., at $z\lesssim 6$, where lensing deflections from large-scale structure are also growing. (For further details on the cosmological evolution of 21cm intensity fluctuations from recombination through reionisation to the present time, see e.g. the review \cite{Pritchard:2011xb}.)} 
Thus  we expect that the lensing contribution to the
bispectrum is nonzero at tree level.

The full-sky redshift-space bispectrum based on \eqref{del2l}--\eqref{eq:DeltaSecOrd} has not been previously presented, as far as we are aware. A partial result was given in \cite{DiDio:2015bua}, where  HI clustering bias and  RSD were neglected in the lensing contribution.  
In \cite{Durrer:2020orn}, the redshift-space bispectrum with HI clustering bias was  presented, but the lensing contribution was omitted.

The article is  structured as follows. In Section \ref{sec:lenContri}, we derive the expression of the  lensing contribution to the bispectrum in  redshift space.
We show  that the lensing contribution is typically much smaller than the unlensed bispectrum. {However,  it can become significant when high-redshift correlations are lensed by a lower redshift fluctuation.}  
We conclude in Section \ref{sec:Conclu}. In  Appendix \ref{sec:trispec}, we present the lensing contribution to the 21cm intensity  4-point correlation function, which is relevant for the variance of the lensed HI intensity mapping power spectrum. 

{In this article, we consider a fiducial flat $\Lambda$CDM cosmology with dimensionless Hubble constant $h=0.67$, baryon and cold dark matter density paramaters $\Omega_\mathrm{b}=0.05$ and  $\Omega_{\mathrm{cdm}}=0.27$, primordial scalar amplitude and tilt $A_\mathrm{s}=2.3\times 10^{-9}$ and   $n_\mathrm{s}=0.962$, evaluated at pivot scale $k_*=0.05\,\mathrm{Mpc}^{-1}$.}

\section{Lensed bispectrum\label{sec:lenContri}}

The  lensed  3-point correlation
function is
\bea
B^{\rm L}(z_{1},\bm{n}_{1},z_{2},\bm{n}_{2},z_{3},\bm{n}_{3})&=&\big\langle\Delta_{1}^{\rm L}\Delta_{2}^{\rm L}\Delta_{3}^{\rm L}\big\rangle
\quad  \text{where }\quad \Delta_{i}\equiv\Delta(z_{i},\bm{n}_{i})
\notag\\
&=&\big\langle\Delta_{1}\Delta_{2}\Delta_{3}\big\rangle+ \delta{B}(z_{1},\bm{n}_{1},z_{2},\bm{n}_{2},z_{3},\bm{n}_{3}).
\eea
At tree level, the lensing correction is 
\begin{equation}
\delta{B}(z_{i},\bm{n}_{i})=\Big\langle\Delta_{1}^{(1)}\Delta_{2}^{(1)}\Big[L_{3}^{(2)}-\big\langle L_{3}^{(2)}\big\rangle\Big]\Big\rangle+2\ \mathrm{perms}.\label{eq:corr_bisp}
\end{equation}
Using \eqref{eq:Lens_Contri} in \eqref{eq:corr_bisp} and applying Wick's
theorem, we find that
\begin{equation}
\delta{B}(z_{i},\bm{n}_{i})={\big\langle\Delta_{1}^{(1)}\, \nabla_{\!\perp}^{a}\phi_{3}\big\rangle\,\big\langle\Delta_{2}^{(1)}\,\nabla_{\!\perp a}\Delta_{3}^{(1)}\big\rangle+\big\langle\Delta_{2}^{(1)}\,\nabla_{\!\perp}^{a}\phi_{3}\big\rangle\,\big\langle\Delta_{1}^{(1)}\,\nabla_{\!\perp a}\Delta_{3}^{(1)}\big\rangle}+2\mathrm{\ perms}.\label{eq:deltaB}
\end{equation}
The corresponding lensing correction to the angular bispectrum is given by
\begin{equation}
{\delta}{B}(z_{i},\bm{n}_{i})=\sum_{\ell_{i}m_{i}}\delta{B}_{\ell_{1}\ell_{2}\ell_{3}}^{m_{1}m_{2}m_{3}}(z_{1},z_{2},z_{3})\,Y_{\ell_{1}m_{1}}(\bm{n}_{1})\,Y_{\ell_{2}m_{2}}(\bm{n}_{2})\,Y_{\ell_{3}m_{3}}(\bm{n}_{3})\,.
\end{equation}

We now derive the expression for ${\delta}{B}_{\ell_{1}\ell_{2}\ell_{3}}^{m_{1}m_{2}m_{3}}(z_{1},z_{2},z_{3})$, starting with the first term of \eqref{eq:deltaB}:
\bea
\big\langle\Delta_{1}^{(1)}\,\nabla_{\!\perp}^{a}\phi_{3}\big\rangle\,\big\langle\Delta_{2}^{(1)}\,\nabla_{\!\perp a}\Delta_{3}^{(1)}\big\rangle 
& =& \sum\big\langle\Delta_{\ell_{1}m_{1}}(z_{1})\,\phi_{\ell_{3}m_{3}}(z_{3})\big\rangle\,\big\langle\Delta_{\ell_{2}m_{2}}(z_{2})\,\Delta_{\ell_{4}m_{4}}(z_{3}) \big\rangle
\nonumber \\
&&~~ \times Y_{\ell_{1}m_{1}}(\bm{n}_{1})\,Y_{\ell_{2}m_{2}}(\bm{n}_{2})\,\nabla_{\!\perp}^{a}\,Y_{\ell_{3}m_{3}}(\bm{n}_{3})\,\nabla_{\!\perp a}\,Y_{\ell_{4}m_{4}}(\bm{n}_{3}).~~~~
\label{eq:corr_calc}
\eea
The harmonic expansion of  gradients of the spherical harmonics can be computed using spin spherical harmonics and the lowering and raising operators 
 \cite{DiDio:2015bua} (see Appendix \ref{app1} for further details). This leads to  
\begin{align}
\nabla_{\!\perp}^{a}Y_{\ell_{3}m_{3}}(\bm{n})\,\nabla_{\!\perp a}Y_{\ell_{4}m_{4}}(\bm{n}) & =-\frac{1}{2} \sqrt{\ell_3 \ell_4 (\ell_3+1)(\ell_4+1)}\,\sum_{\ell m} (-1)^{m}\,Y_{\ell m}(\bm{n})\Big[1+(-1)^{\ell_3+\ell_4+\ell}\Big] \nonumber \\
 & \times \sqrt{\frac{(2\ell+1)(2\ell_3+1)(2\ell_4+1)}{4 \pi}}\begin{pmatrix}\ell_{3} & \ell_{4} & \ell\\
m_{3} & m_{4} & -m
\end{pmatrix}\begin{pmatrix}\ell_{3} & \ell_{4} & \ell\\
1 & -1 & 0
\end{pmatrix},
\label{eq:Deri_Dot}
\end{align}
where the $3\times 2$ matrices  are Wigner 3j symbols {(evaluated with the \texttt{wigxjpf} code \cite{Johansson:2015cca})}.  The second term of \eqref{eq:deltaB} follows similarly.

Using \eqref{eq:corr_calc}   and  \eqref{eq:Deri_Dot}, together with their counterparts for the second term of \eqref{eq:deltaB},  we find that the lensing contribution to the angular bispectrum is
\begin{align}
\delta{B}_{\ell_{1}\ell_{2}\ell_{3}}^{m_{1}m_{2}m_{3}} & =-\left[C_{\ell_{1}}^{\Delta\Delta}(z_{1},z_{3})\,C_{\ell_{2}}^{\Delta\phi}(z_{2},z_{3})+C_{\ell_{1}}^{\Delta\phi}(z_{1},z_{3})\,C_{\ell_{2}}^{\Delta\Delta}(z_{2},z_{3})\right]\!\begin{pmatrix}\ell_{1} & \ell_{2} & \ell_{3}\\
1 & -1 & 0
\end{pmatrix}\!\begin{pmatrix}\ell_{1} & \ell_{2} & \ell_{3}\\
m_{1} & m_{2} & m_{3}
\end{pmatrix}\nonumber \\
 & \times\sqrt{\frac{\ell_{1}\ell_{2}(\ell_{1}+1)(\ell_{2}+1)(2\ell_{1}+1)(2\ell_{2}+1)(2\ell_{3}+1)}{4\pi}}+2\mathrm{\ perms}.\label{eq:bisp_m_ell}
\end{align}
{Here the first-order angular power spectra $C_\ell^{XY}$ are defined by}
\begin{equation}
\left\langle X_{\ell m}(z)\,Y_{\ell'm'}(z')\right\rangle =(-1)^{m}\,C_{\ell}^{XY}(z,z')\,\delta_{\ell\ell'}\,\delta_{m,-m'}\,,\label{eq:pow-spec-def}
\end{equation}
{where $\Delta$ denotes $\Delta_{\rm HI}^{(1)}$ and $\phi$ denotes $\phi^{(1)}$, so that $C^{\Delta\Delta}_\ell$ is the HI intensity auto power spectrum  and $C^{\Delta\phi}_\ell$ is the cross power spectrum of the lensing potential with HI intensity.}  
 
Statistical isotropy allows us to define the reduced lensing contribution to the bispectrum: 
\begin{equation}
{\delta}B_{\ell_{1}\ell_{2}\ell_{3}}^{m_{1}m_{2}m_{3}}(z_1,z_2,z_3)=\mathcal{G}_{\ell_{1}\ell_{2}\ell_{3}}^{m_{1}m_{2}m_{3}}\, {\delta}b_{\ell_{1}\ell_{2}\ell_{3}}(z_1,z_2,z_3)\,,\label{eq:red-bisp-def}
\end{equation}
where  the Gaunt integral is
\begin{equation}
\mathcal{G}_{\ell_{1}\ell_{2}\ell_{3}}^{m_{1}m_{2}m_{3}}=\sqrt{\frac{(2\ell_{1}+1)(2\ell_{2}+1)(2\ell_{3}+1)}{4\pi}}\begin{pmatrix}\ell_{1} & \ell_{2} & \ell_{3}\\
m_{1} & m_{2} & m_{3}
\end{pmatrix}\begin{pmatrix}\ell_{1} & \ell_{2} & \ell_{3}\\
0 & 0 & 0
\end{pmatrix}.
\end{equation}
\begin{figure}[!h]
\begin{centering}
\includegraphics[scale=0.5]{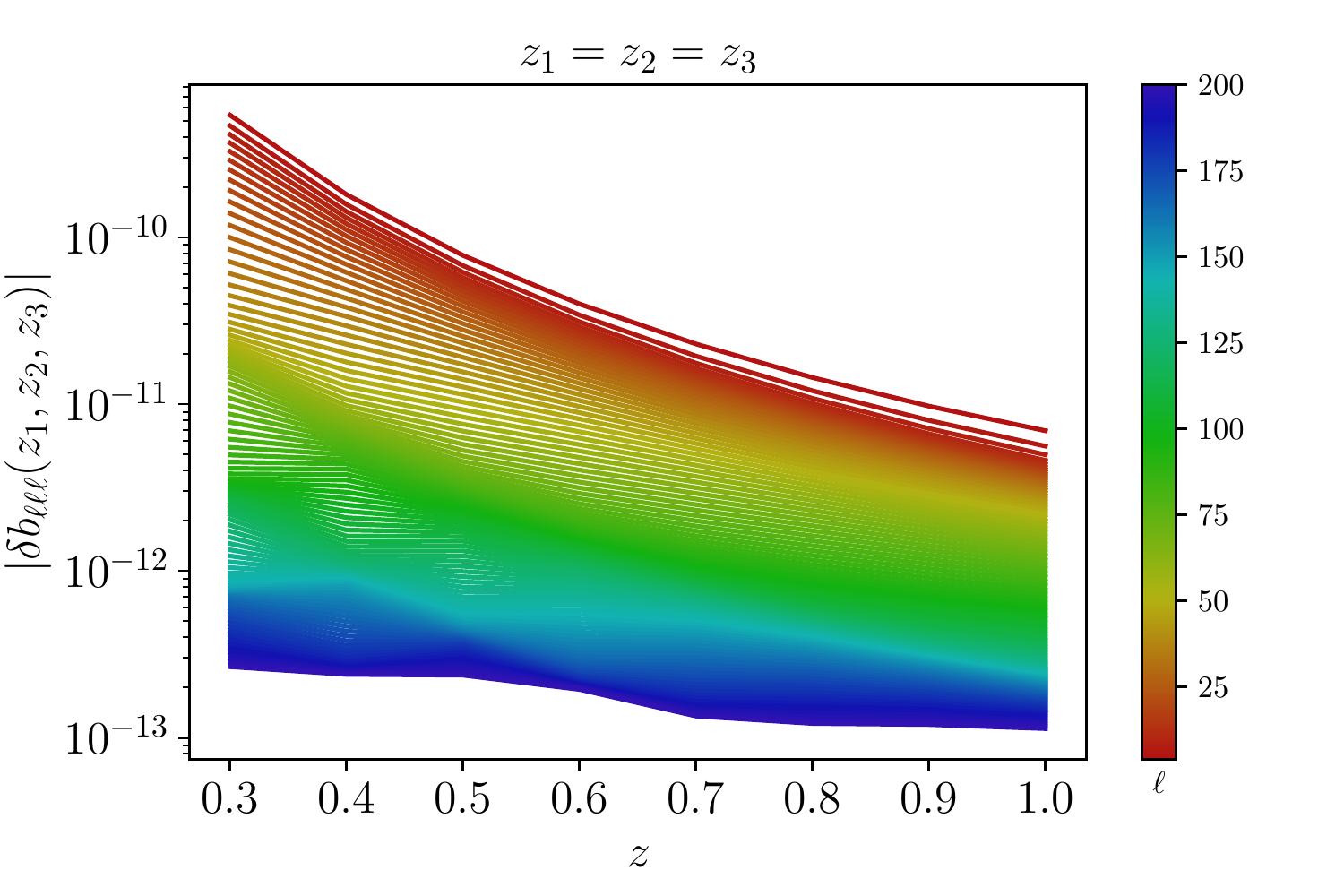}
\includegraphics[scale=0.5]{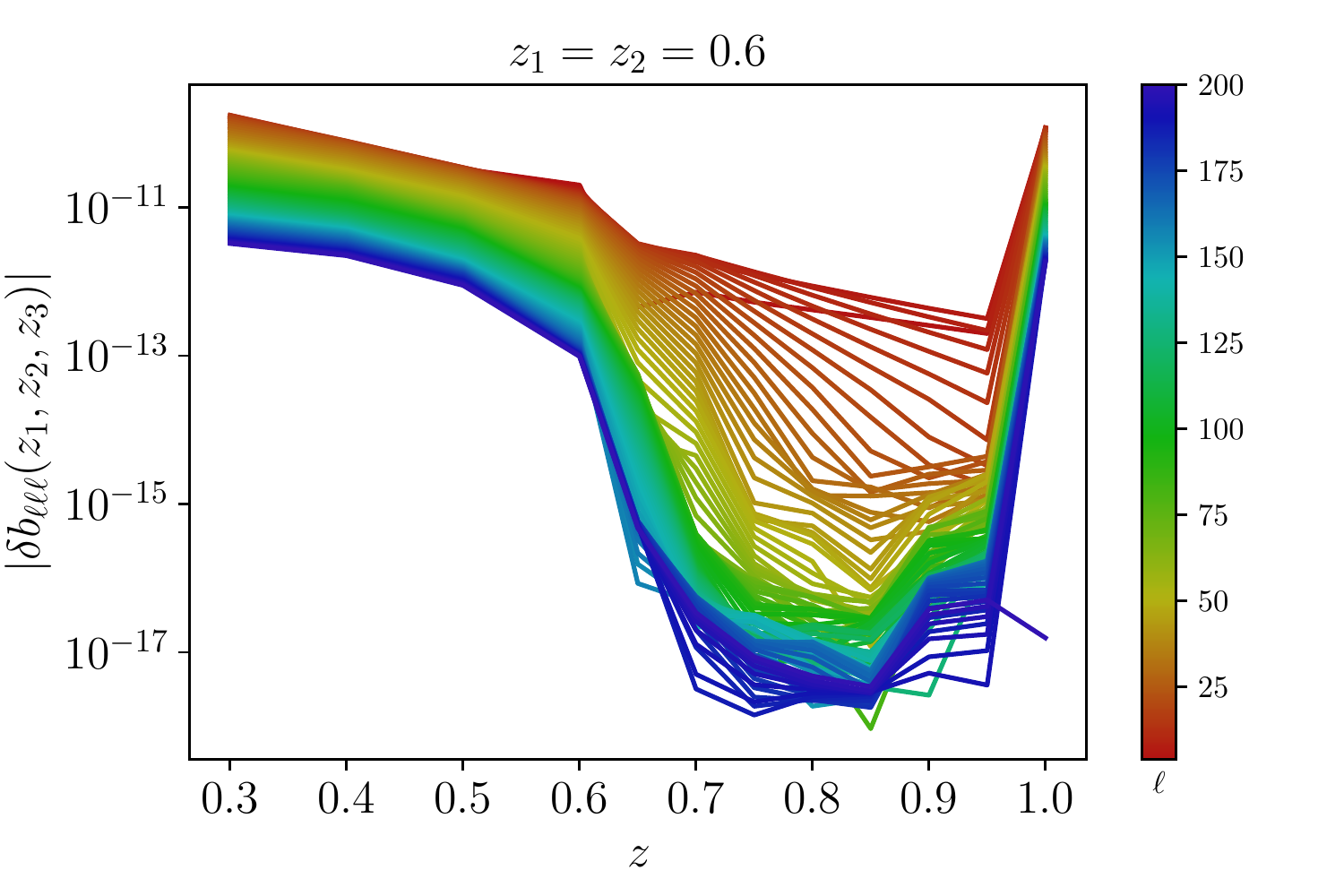}
\par\end{centering}
\caption{Lensing contribution ${\delta}b_{\ell\ell\ell}$ to the 21cm intensity bispectrum in the equilateral configuration, with  three equal redshifts (left) and $z_1=z_2=0.6$ fixed with  varying $z_3$ (right). Colour bar shows the $\ell$ value. 
\label{fig:Bisp_Cont}}
\end{figure}
\begin{figure}[!h]
\begin{centering}
\includegraphics[scale=0.5]{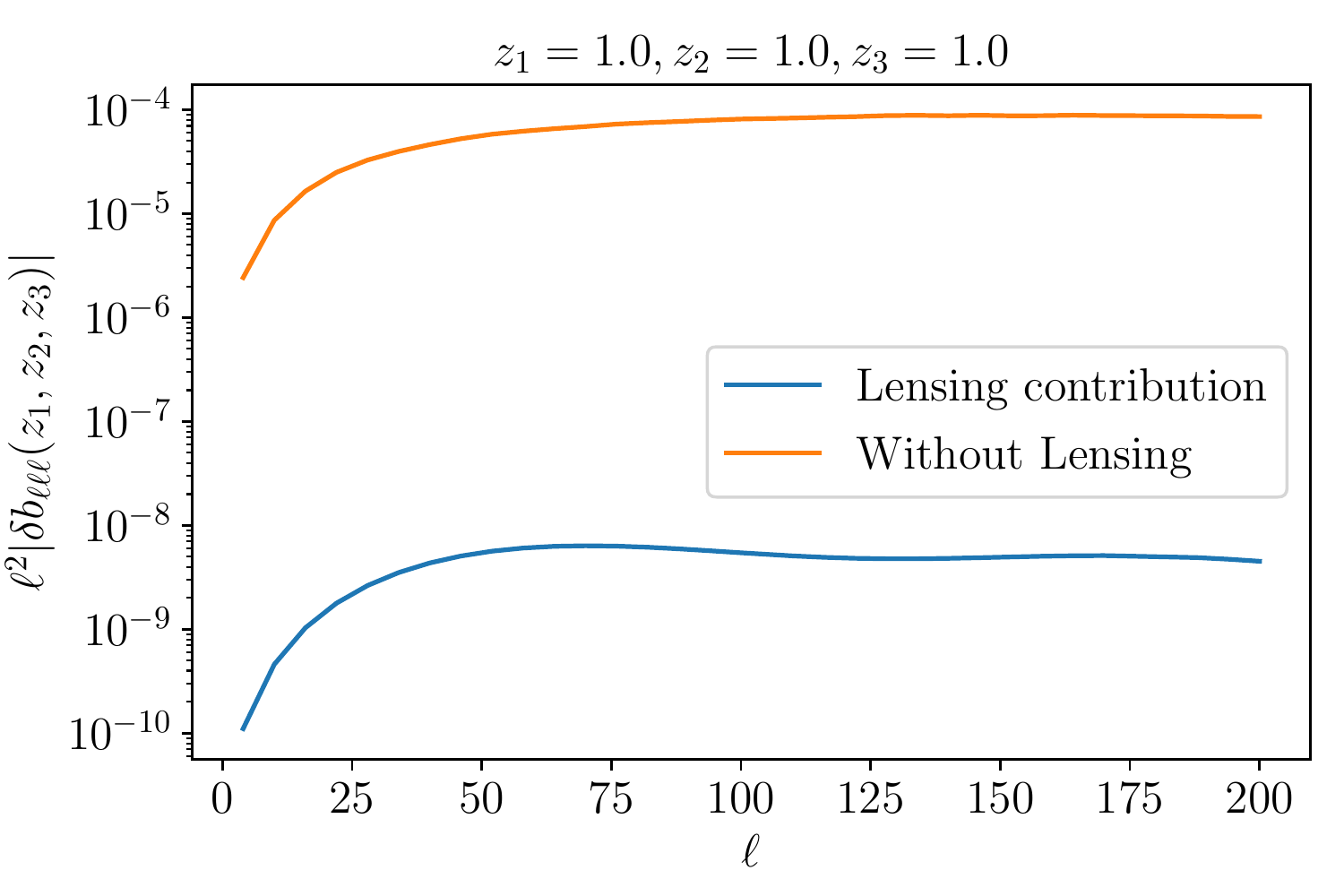}
\includegraphics[scale=0.5]{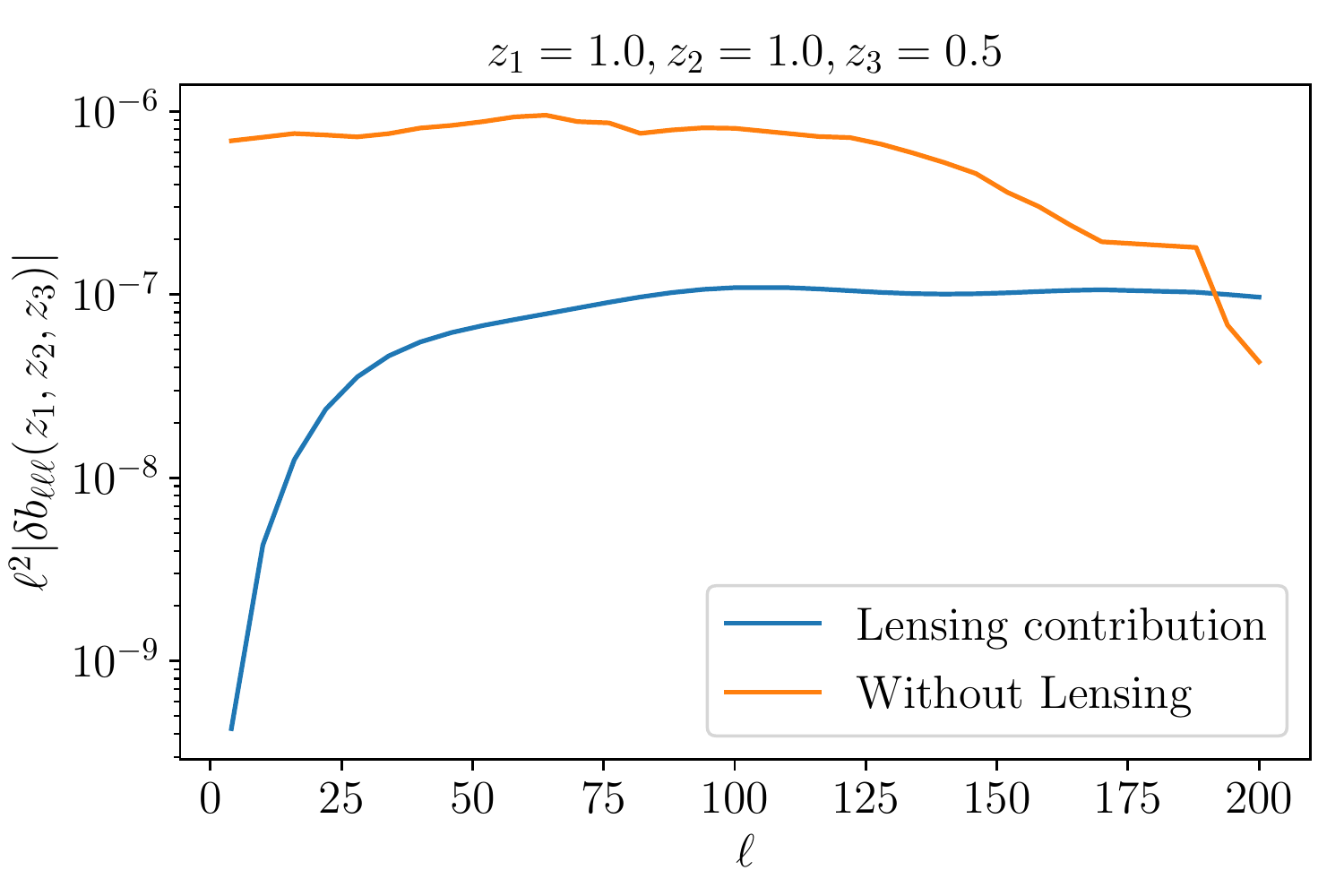}\\
\includegraphics[scale=0.5]{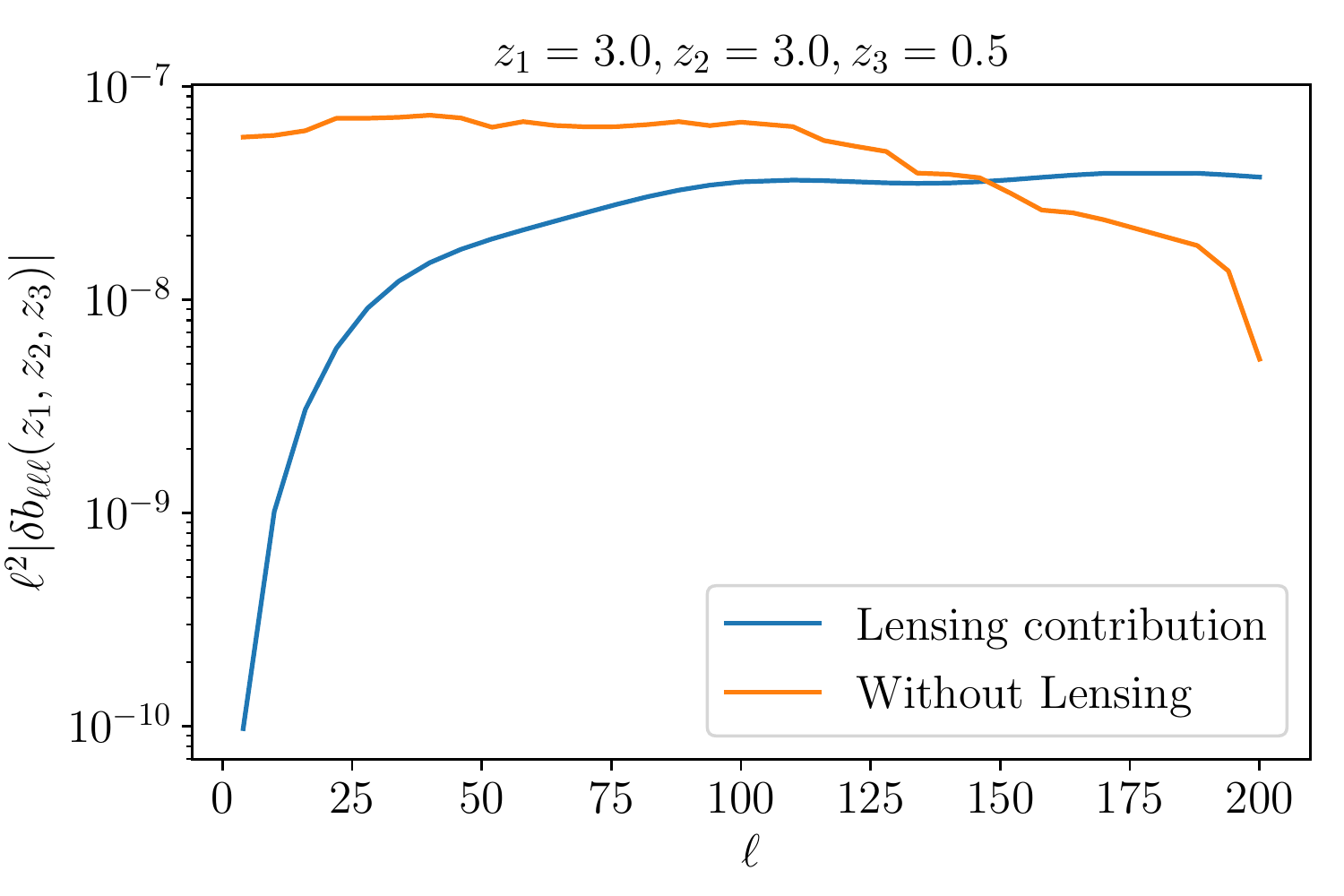}
\includegraphics[scale=0.5]{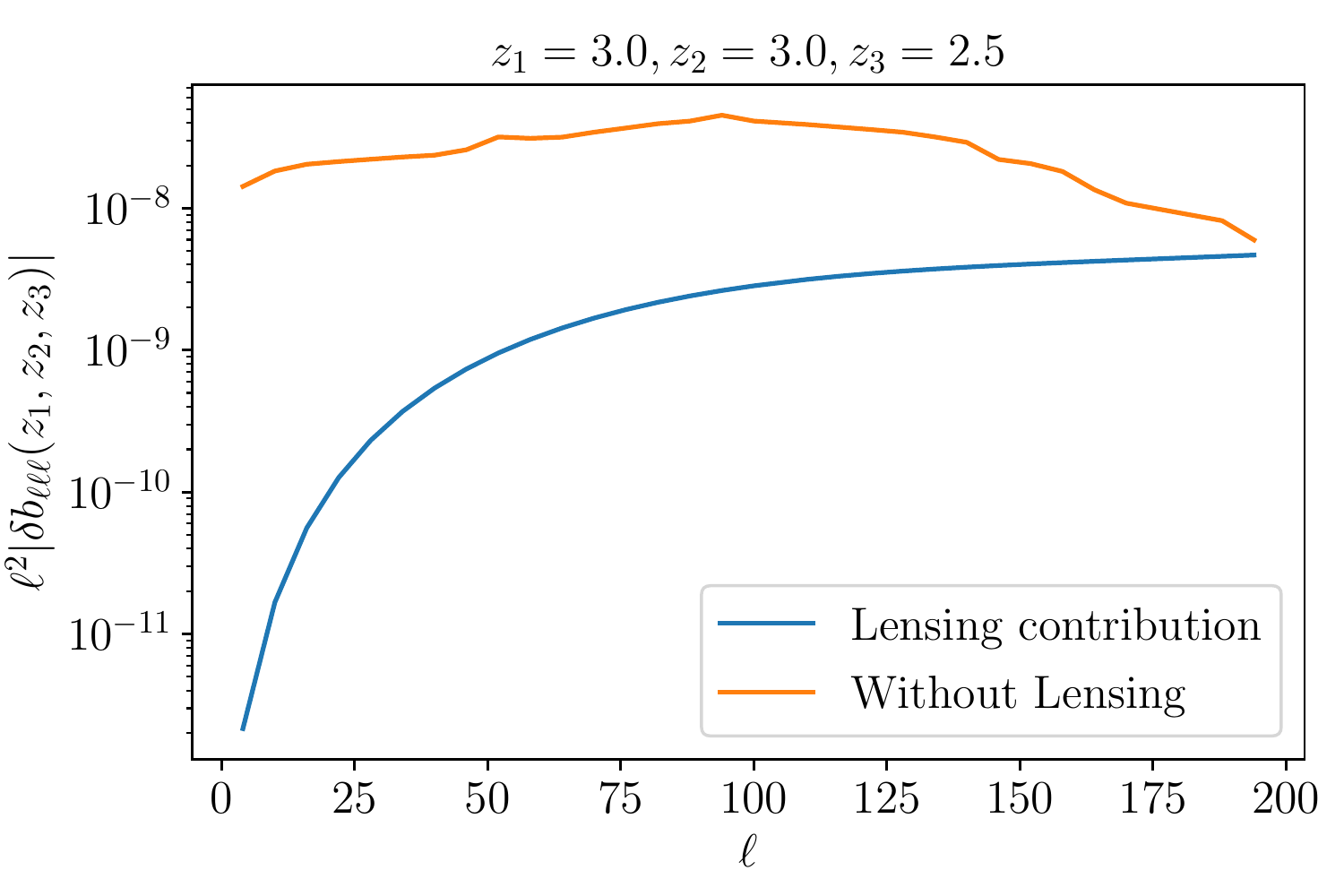}
\par\end{centering}
\caption{21cm intensity mapping bispectrum in the equilateral configuration and for various redshift triples: without lensing ($b_{\ell\ell\ell}$, orange),  lensing contribution (${\delta}b_{\ell\ell\ell}$, blue). \label{fig:HI_Len_Comp}}
\end{figure}
\begin{figure}[!h]
\begin{centering}
\includegraphics[scale=0.5]{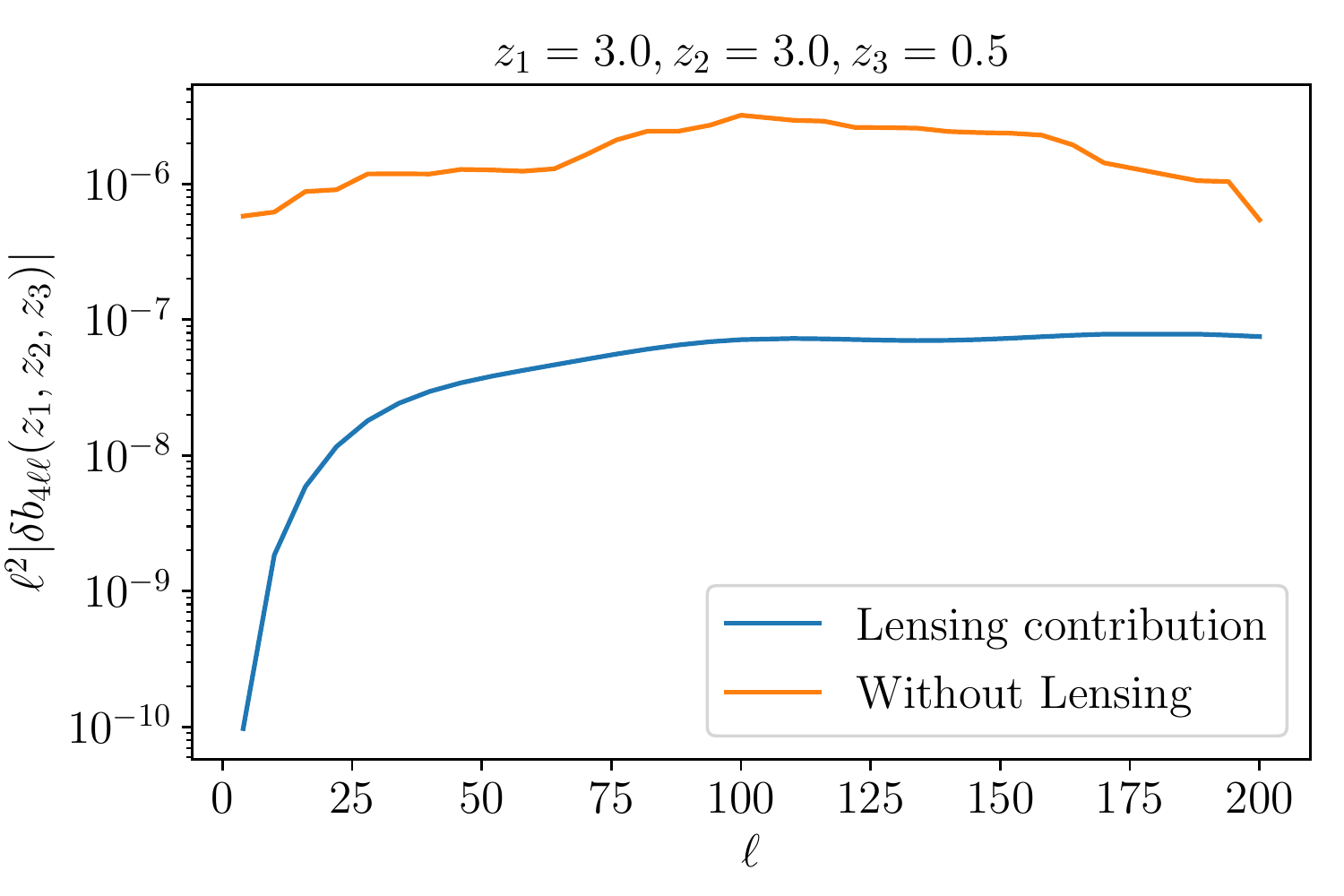}
\includegraphics[scale=0.5]{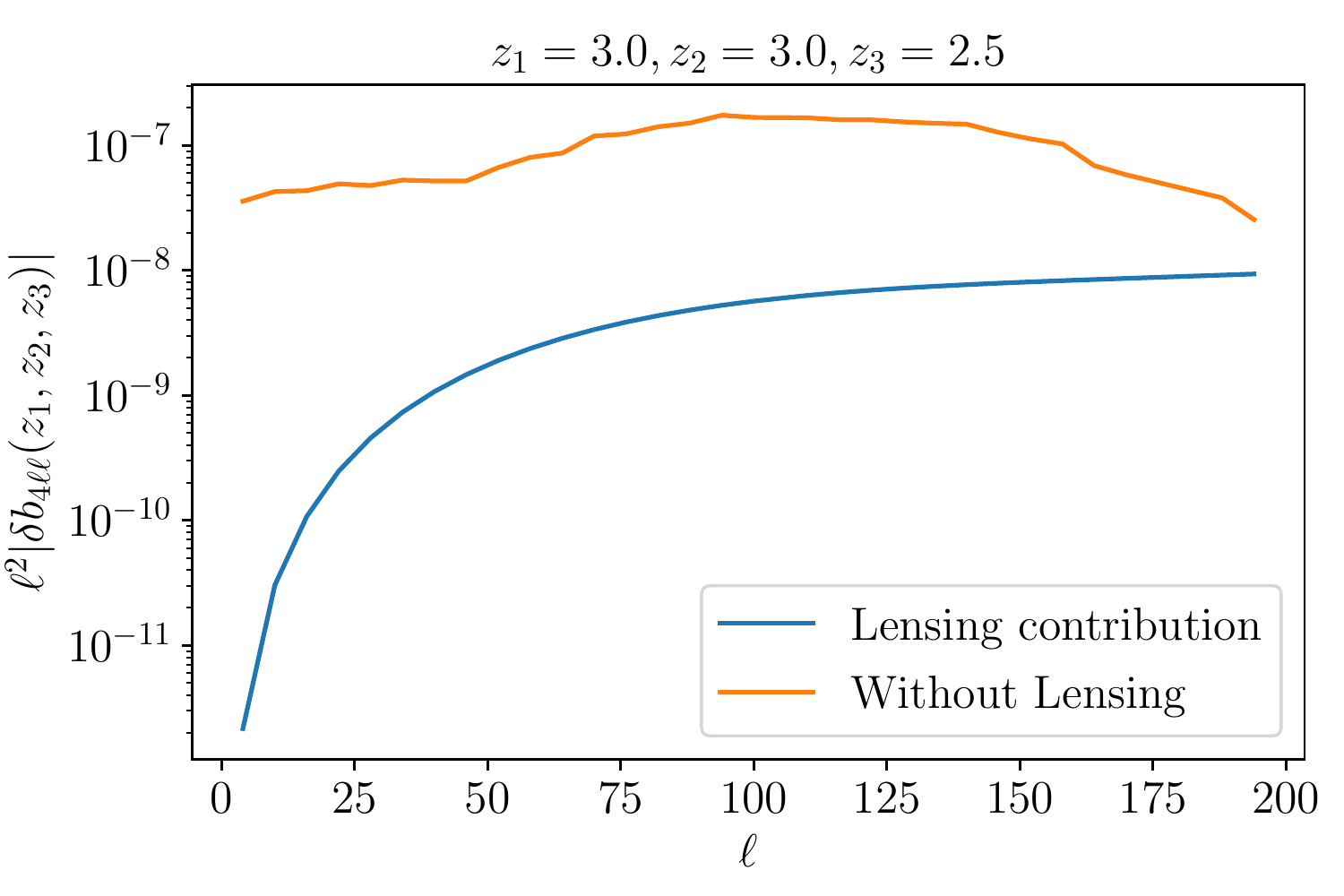}
\par\end{centering}
\caption{As in Figure \ref{fig:HI_Len_Comp},  for isosceles configurations with $\ell_1=4, \ell_2=\ell_3\equiv\ell$. Squeezed configurations have $\ell \gg 4$. 
\label{squeezed}}
\end{figure}
From \eqref{eq:bisp_m_ell} and \eqref{eq:red-bisp-def}, it follows
that the lensing contribution to the reduced bispectrum is given by
\begin{align}
{\delta}b_{\ell_{1}\ell_{2}\ell_{3}} & =-\left[C_{\ell_{1}}^{\Delta\Delta}(z_{1},z_{3})\,C_{\ell_{2}}^{\Delta\phi}(z_{2},z_{3})+C_{\ell_{1}}^{\Delta\phi}(z_{1},z_{3})\,C_{\ell_{2}}^{\Delta\Delta}(z_{2},z_{3})\right]\nonumber \\
 & \times \begin{pmatrix}\ell_{1} & \ell_{2} & \ell_{3}\\
0 & 0 & 0
\end{pmatrix}^{-1} \begin{pmatrix}\ell_{1} & \ell_{2} & \ell_{3}\\
1 & -1 & 0
\end{pmatrix}\sqrt{\ell_{1}\ell_{2}(\ell_1+1)(\ell_2+1)}+2\mathrm{\ perms}.\label{eq:RedBispExp}
\end{align}
This is our main result. It extends the result of \cite{Durrer:2020orn}, which presented and computed the unlensed $b_{\ell_{1}\ell_{2}\ell_{3}}$ for 21cm intensity maps,  and it recovers the special case
in \cite{DiDio:2015bua}, where the RSD and clustering bias effects were neglected in ${\delta}b_{\ell_{1}\ell_{2}\ell_{3}}$.  

Examples of the absolute value of the reduced bispectrum \eqref{eq:RedBispExp}  are shown in Figures \ref{fig:Bisp_Cont}--\ref{squeezed}. We used \texttt{CLASS} \cite{Blas:2011rf,Lesgourgues:2011re} for the lensed contribution ${\delta}b_{\ell_{1}\ell_{2}\ell_{3}}$ and the \textsc{Byspectrum} code\footnote{https://gitlab.com/montanari/byspectrum}   \cite{DiDio:2018unb,Durrer:2020orn} for the unlensed bispectrum $b_{\ell_{1}\ell_{2}\ell_{3}}$. Following \cite{Durrer:2020orn}, we modelled the   HI clustering bias parameters as
\bea
b_{1}(z) &=&~~0.754 +0.0877z +0.0607z^{2} -0.00274z^{3}\,,
\label{b1}\label{eq:bias_b1}\\ 
b_{2}(z)	 &=& -0.308-0.0724z -0.0534z^{2}+ 0.0247z^{3}\,, 
\label{b2}\\
b_{s}(z)	&=&-\frac{2}{7}\big[b_{1}(z)-1\big].
\label{bs}
\eea 
Here $b_1, b_2$ are cubic fits to halo model predictions, while $b_s$ is the simplest tidal bias model, corresponding to vanishing initial tidal bias.

Figure \ref{fig:Bisp_Cont}  displays the lensing contribution to the reduced bispectrum in the equilateral configuration, colour-coded according to the multipole values $\ell$, with all three redshifts the same  (left) and  with $z_1=z_2=0.6$ and varying $z_3$ (right). The left panel shows that the lensing contribution in the equal-redshift case decreases as $z$ and $\ell$ increase. 
The right panel shows that for two equal redshifts, the signal is greater when the third redshift is smaller -- i.e., when the equal-redshift fluctuations are lensed by the lower redshift fluctuation. This is consistent with examples for galaxy surveys given in \cite{DiDio:2015bua}.

Figure \ref{fig:HI_Len_Comp} compares  the lensing contribution to the unlensed reduced bispectrum in the equilateral configuration for various redshift triples.
Appropriate smoothing of the unlensed bispectrum with a {15-point average} filter has been performed where necessary {(see \cite{DiDio:2018unb,Durrer:2020orn} for discussion of numerical issues in the redshift-space angular bispectrum)}. 
The bottom panels show a striking example of how the relative lensing contribution peaks when  equal-redshift fluctuations are lensed by a lower redshift fluctuation. With high equal redshifts ($z\sim3$), the lensing contribution can become comparable to, or even dominate over, the density and RSD contributions.

In Figure \ref{squeezed}, the lensing contribution in isosceles configurations, $ \ell_2=\ell_3\equiv\ell$ with $\ell_1=4$, is illustrated for the same redshift triples as the bottom row of Figure \ref{fig:HI_Len_Comp}. For $\ell \gg 4$, we approach the squeezed limit. This case shows a similar behaviour to the equilateral, although the relative lensing contribution is higher in the equilateral case.

\section{Discussion and conclusion\label{sec:Conclu}}

We derived the lensing contribution to the full-sky HI intensity mapping bispectrum in redshift space, at tree level, as given in \eqref{eq:RedBispExp}. This generalises earlier results to include all RSD effects, as well as the clustering bias up to second order (including tidal bias). We presented some numerical examples for the equilateral configuration in Figures \ref{fig:Bisp_Cont} and \ref{fig:HI_Len_Comp}, and for the isosceles (including squeezed) configuration in Figure \ref{squeezed}. 

These examples suggest that the lensing contribution is greatest when two equal-redshift fluctuations are lensed by a lower redshift fluctuation, as expected from previous work on the galaxy bispectrum. For example, in the equilateral case with $z_1=z_2=3$, $z_3=0.5$, the lensing contribution dominates in amplitude over the density and RSD contributions for $\ell \gtrsim 100$.

For other redshift configurations, including three equal redshifts, the lensing contribution is orders of magnitude below the unlensed contribution. Equation \eqref{eq:RedBispExp}, shows that the 21cm lensing arises from terms of the form $C_{\ell}^{\Delta\Delta}\,C_{\ell'}^{\Delta\phi}$. The lensing contribution is only in 
$C_{\ell'}^{\Delta\phi}$, and for equal redshifts
$C_{\ell}^{\Delta\Delta}\gg C_{\ell'}^{\Delta\phi}$ {(for equal redshifts $C_{\ell}^{\Delta\phi}>0$)}. As a consequence, the lensing effect is swamped by the contributions of density and RSD. A way out of this is if the redshifts are unequal, when it is possible (as shown in our examples) that the contribution of $C_{\ell}^{\Delta\Delta}$ is heavily suppressed while the lensing deflection can lead to an enhanced $C_{\ell'}^{\Delta\phi}$.

\begin{figure}[!!h]
\begin{centering}
\includegraphics[scale=0.35]{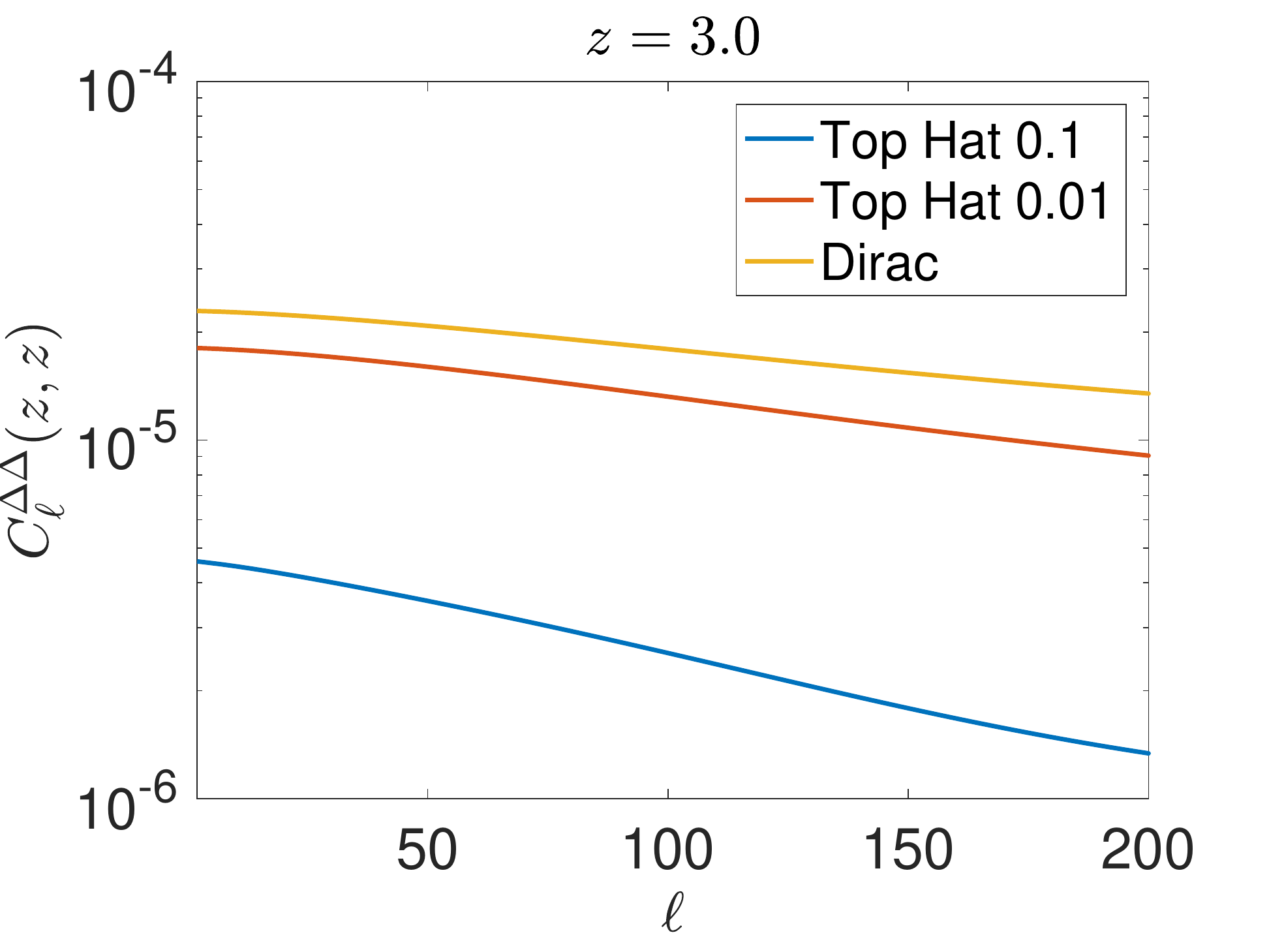}
\includegraphics[scale=0.35]{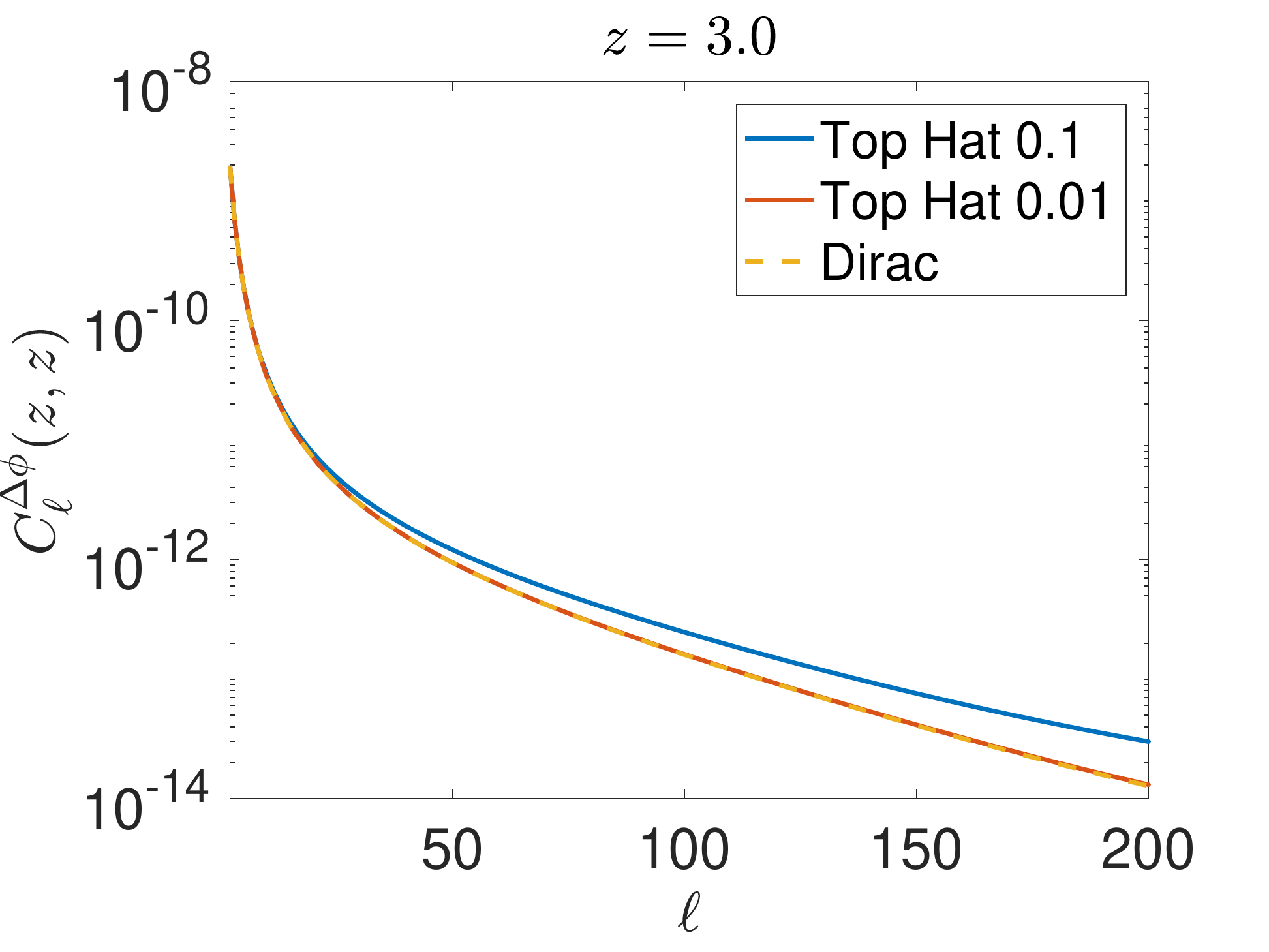}
\includegraphics[scale=0.35]{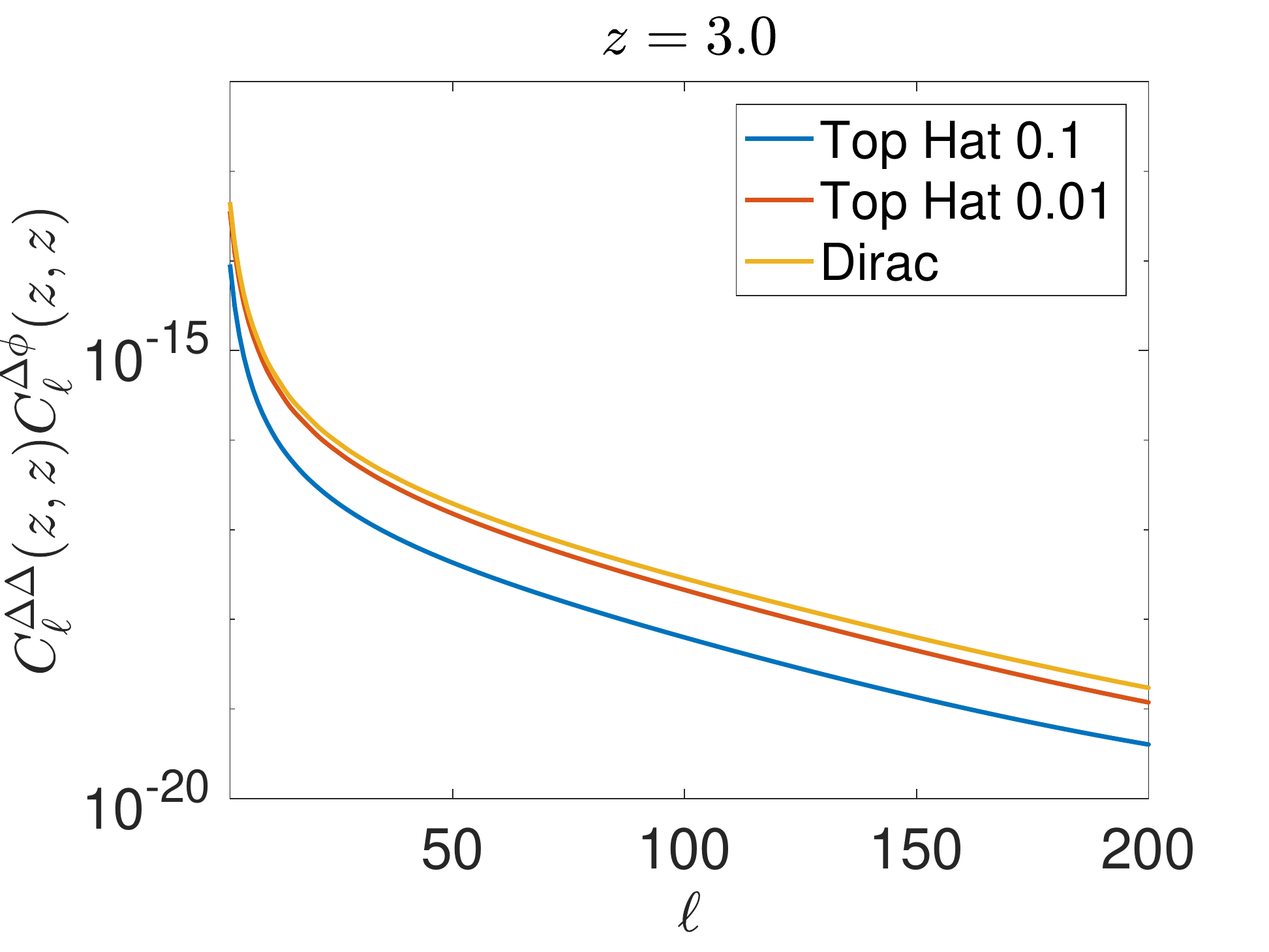}\\
\par\end{centering}
\caption{Effect of varying the $z$-bin width, using a top-hat window function, for an equilateral configuration and equal redshifts. {\em Top:} The 2 angular power spectra in the lensing reduced bispectrum \eqref{eq:SpecCaseA}. {\em Bottom:} The product of the power spectra.
\label{zbins}}
\end{figure}
\begin{figure}[!h]
\begin{centering}
\includegraphics[scale=0.35]{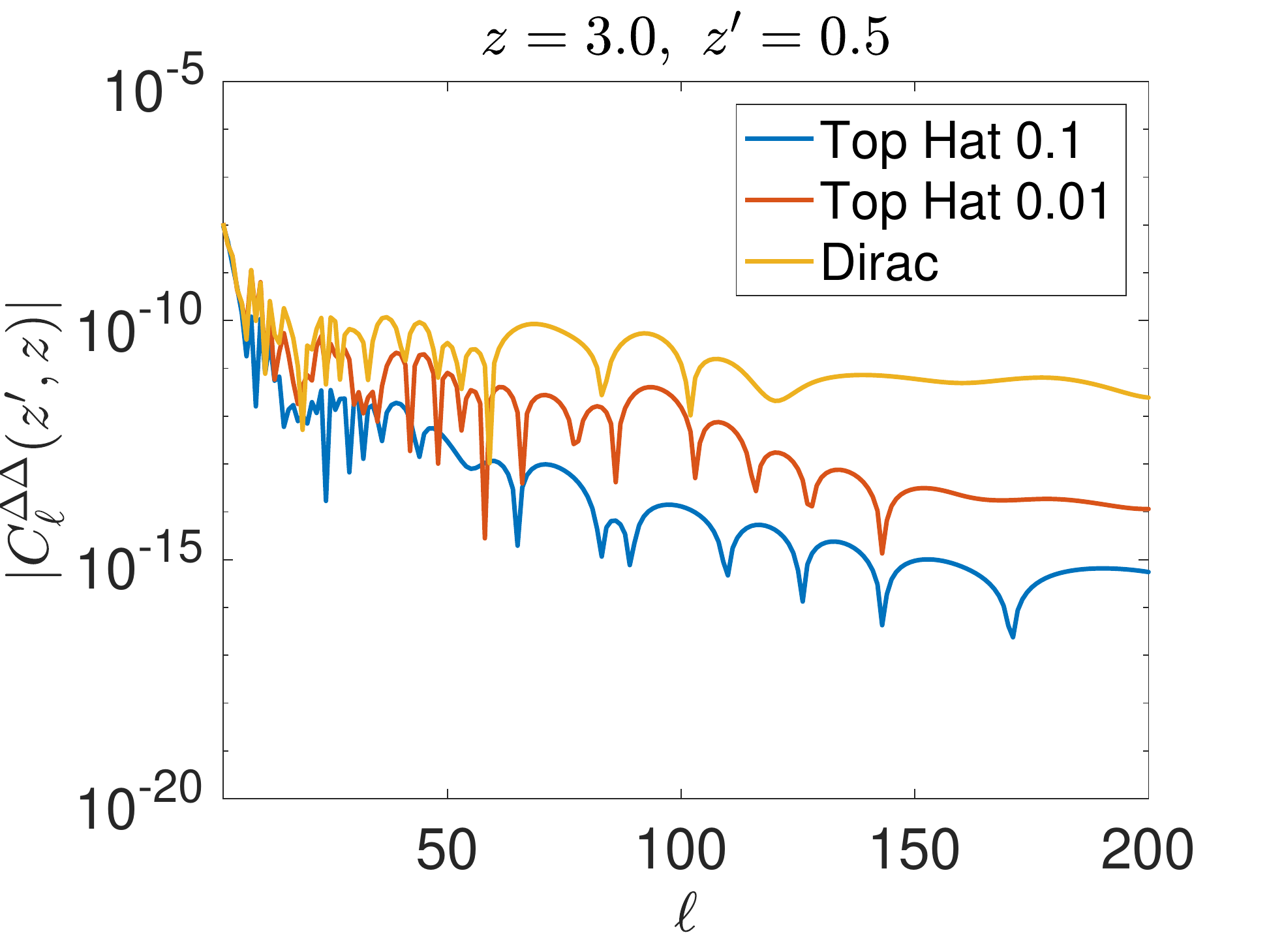}\includegraphics[scale=0.35]{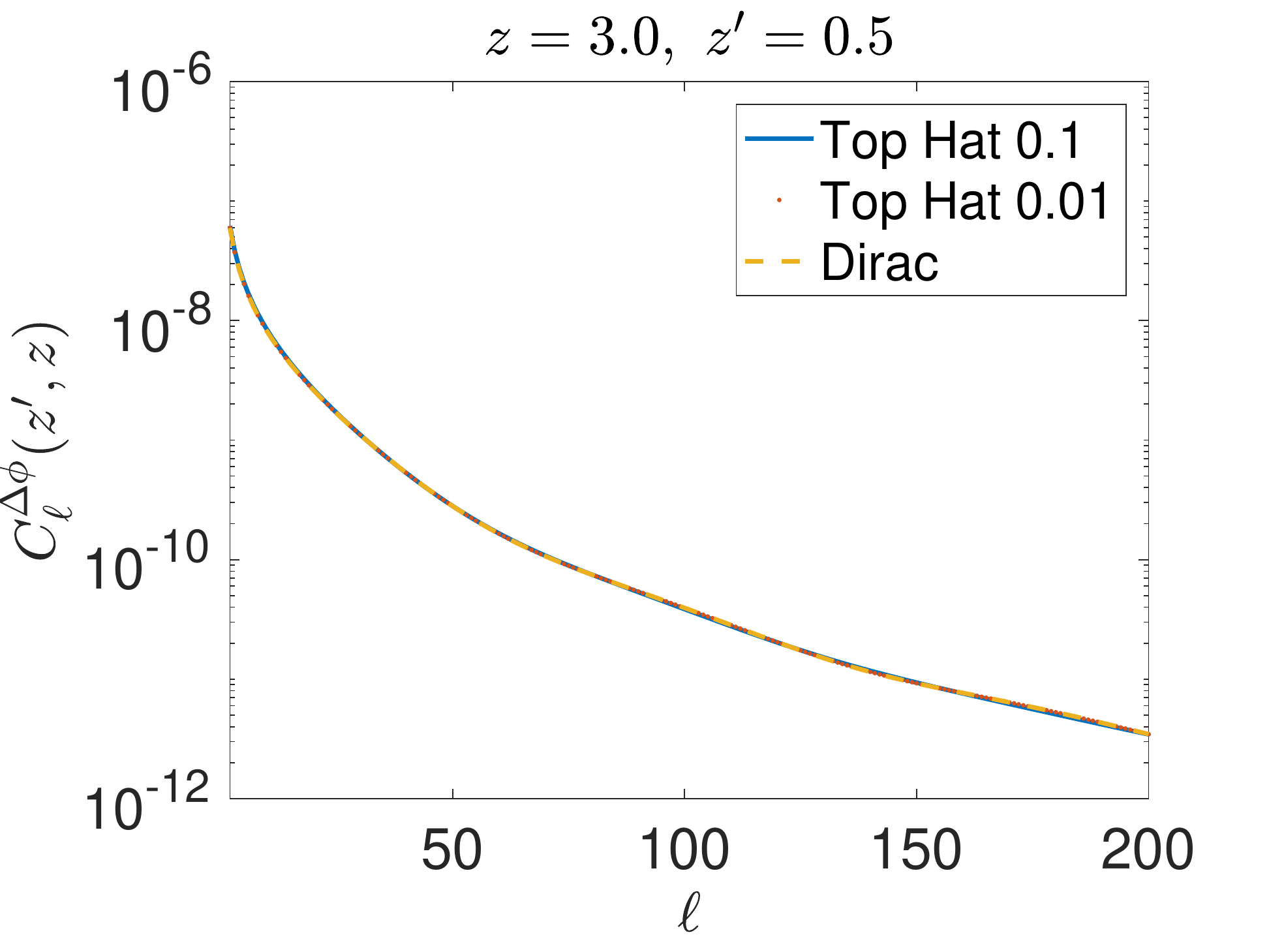}
\includegraphics[scale=0.35]{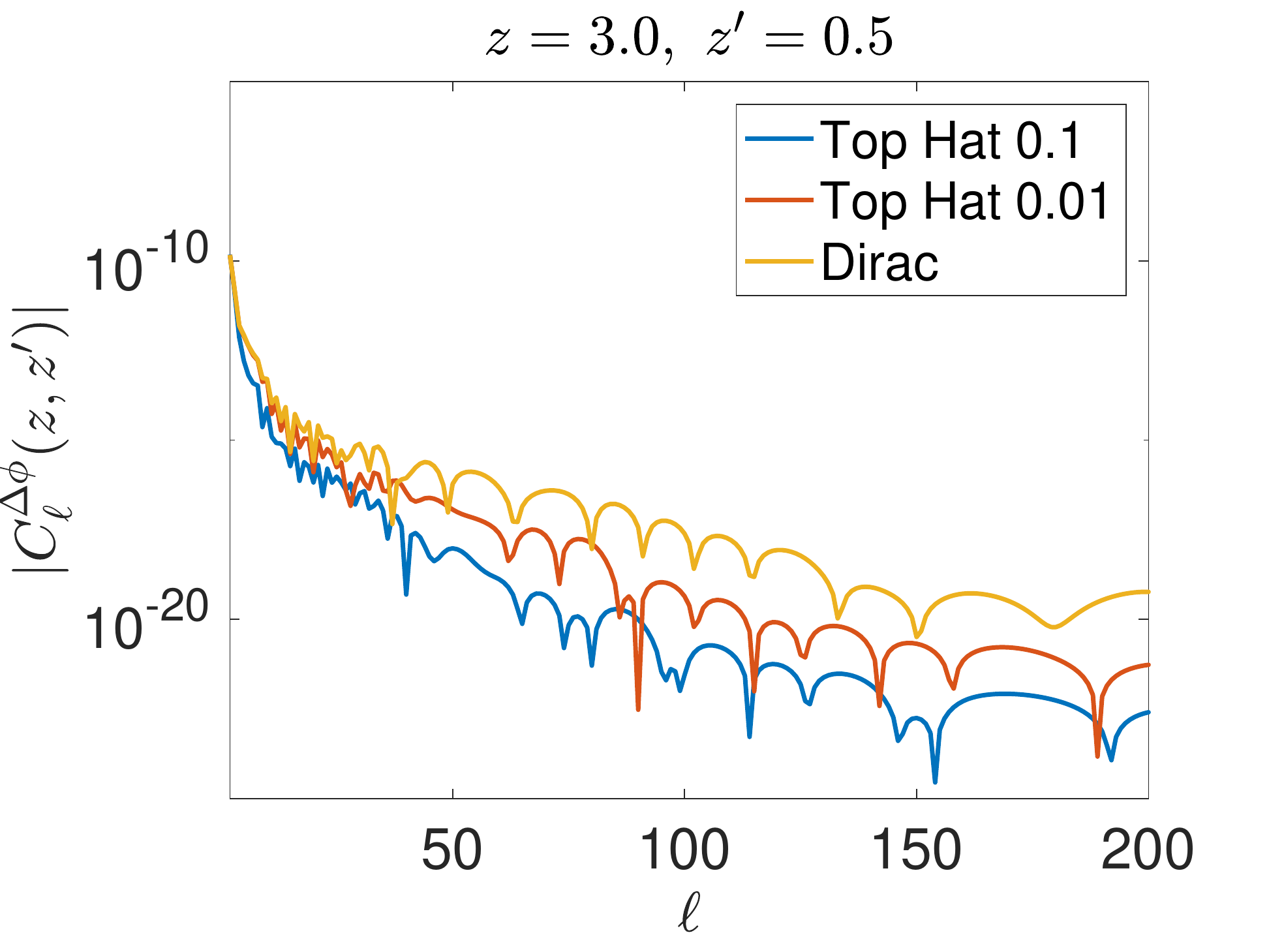}\includegraphics[scale=0.35]{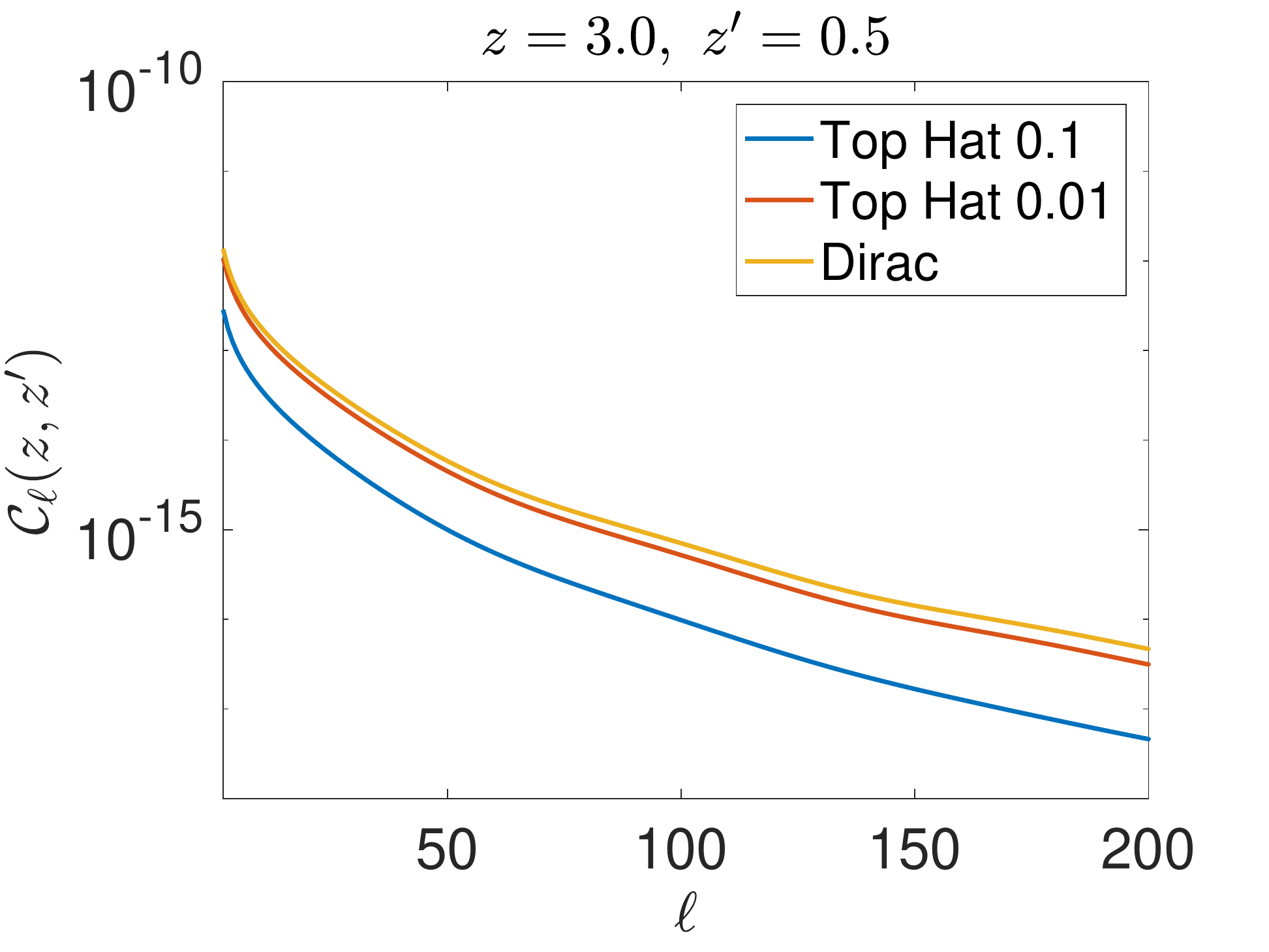}
\par\end{centering}
\caption{
As in Figure \ref{zbins}, but with $z_1=z_2\equiv z=3, z_3\equiv z'=0.5$. The angular power spectra in the lensing reduced bispectrum \eqref{eq:SpecCaseB}, in addition to the 2 in Figure \ref{zbins}, are shown in the top and bottom left panels. The bottom right panel shows $\bm{\mathcal{C}}_\ell$, defined
 in \eqref{eq:SpecCaseB}.}
\label{zbins2}
\end{figure}

The relative 21cm lensing effect is typically much smaller than in galaxy surveys, for two reasons. 
\begin{itemize}
    \item 
There is a first-order lensing contribution for galaxies, proportional to the lensing convergence $\kappa^{(1)}$ \cite{Alonso:2015uua}, which is absent from  21cm intensity. 
\item
At second order, the lensing effect in the  21cm bispectrum is from a single contribution,  $L^{(2)}=\nabla_{\!\perp}^{a}\phi^{(1)}\,\nabla_{\!\perp a}\Delta^{(1)}$, whereas for galaxies, there are many more  contributions, from $\kappa^{(2)}$, $[\kappa^{(1)}]^2$, $\kappa^{(1)}\Delta^{(1)}$, $\nabla_{\!\perp}^{a}\kappa^{(1)}\,\nabla_{\!\perp a}\phi^{(1)}$ (see \cite{DiDio:2015bua}). It follows that the lensing effect for galaxies has pure-lensing and lensing $\times$ (density + RSD) contributions, whereas for 21cm intensity, we have only a single lensing $\times$ (density + RSD) contribution.   
\end{itemize}

In our numerical examples we have used infinitely thin redshift bins. In the case of the galaxy angular power spectrum, it is known that increasing the width of redshift bins suppresses the density and RSD signals, but  can increase the lensing contribution \cite{DiDio:2013sea,Cardona:2016qxn,Jelic-Cizmek:2020pkh}. Similar behaviour is seen in the galaxy angular bispectrum  \cite{DiDio:2015bua}.
By contrast, in the 21cm bispectrum  the lensing contribution is typically {\em suppressed} by increasing the bin width.  

The reason for this is again rooted in the fact that 21cm lensing is sourced by terms of the form $C_{\ell}^{\Delta\Delta}\,C_{\ell'}^{\Delta\phi}$. Although the cross-power spectrum with lensing, $C_{\ell'}^{\Delta\phi}$, may be enhanced by increasing the bin width, it is always multiplied by a non-lensing auto-power spectrum $C_{\ell}^{\Delta\Delta}$, which is suppressed by increasing the bin width.

This is illustrated by
using redshift bins of  width  $\delta z= 0$ (Dirac delta window), 0.1 and 0.01, with a top-hat window function. 
We compute two equilateral examples as follows.

\noindent \underline{$z_i=z=3$:}\quad  The reduced lensing bispectrum \eqref{eq:RedBispExp} is
\bea
\delta b_{\ell\ell\ell}(z,z,z)&=&3\alpha_\ell \, C_{\ell}^{\Delta\Delta}(z,z)\,C_{\ell}^{\Delta\phi}(z,z)\,,\label{eq:SpecCaseA}\\
\alpha_\ell &\equiv&-2\ell(\ell+1)\begin{pmatrix}\ell & \ell & \ell\\
0 & 0 & 0
\end{pmatrix}^{-1}\begin{pmatrix}\ell & \ell & \ell\\
1 & -1 & 0
\end{pmatrix}\,.
\eea
Figure \ref{zbins} shows the 2 power spectra in \eqref{eq:SpecCaseA} (top panels) and then their product (bottom panel), which is proportional to the lensing contribution to the bispectrum.  It can be seen that although the cross-power spectrum of lensing with HI intensity increases with bin width (top right panel), this contribution is overpowered by the effect of the HI auto-power spectrum, which decreases with bin  width (top left). The lensing contribution to the bispectrum thus decreases with bin width, as follows from the bottom panel and \eqref{eq:SpecCaseA}.\\ 

\noindent \underline{$z_1=z_2\equiv z=3, z_3\equiv z'=0.5$:}\quad 
The reduced lensing bispectrum \eqref{eq:RedBispExp} is
\bea
\delta b_{\ell\ell\ell}(z,z,z^{\prime}) &=& \alpha_\ell\,\bm{\mathcal{C}}_{\ell}(z,z')\,,\\
\bm{\mathcal{C}}_{\ell}(z,z')&\equiv & C_{\ell}^{\Delta\Delta}(z,z')\,C_{\ell}^{\Delta\phi}(z,z')+C_{\ell}^{\Delta\Delta}(z',z)\,C_{\ell}^{\Delta\phi}(z,z)+C_{\ell}^{\Delta\Delta}(z,z)\,C_{\ell}^{\Delta\phi}(z',z) .~~~~
\label{eq:SpecCaseB}
\eea
In this case, the two angular power spectra of \eqref{eq:SpecCaseA} are included in \eqref{eq:SpecCaseB}, together with three further power spectra, noting that $C_{\ell}^{\Delta\Delta}(z',z)=C_{\ell}^{\Delta\Delta}(z,z')$. These three additional power spectra appearing in \eqref{eq:SpecCaseB} are shown in
Figure \ref{zbins2} (top panels and bottom left panel). The bottom left panel hows an example of a lensing contribution that decreases with
bin width. Once again the total lensing contribution decreases with bin width, as follows from the bottom right panel and \eqref{eq:SpecCaseB}.

We have only considered a single redshift triple in our examples. In practice, the correlations from many triples will be added and this may enhance the lensing contribution. 
{The 21cm bispectrum has been shown to be detectable by SKA (Phase 1) and HIRAX in \cite{Durrer:2020orn}, and therefore it could be measured using standard estimators in the literature. Detectability of the lensing contribution requires significant further work. Our initial rough estimates indicate that the signal to noise ratio of the lensing signal 
in equal-redshift bins for  SKA1 and HIRAX is small and cross-bin correlations will need to be included for the possibility of a future  detection.}\\

\clearpage

\noindent{\bf Acknowledgements:}
We thank Ruth Durrer, Mona Jalilvand and Francesco Montanari for helpful comments, especially on Appendix  \ref{sec:trispec}. RK thanks Saurabh Kumar for help with coding.
RK and RM are supported by the South African Radio Astronomy Observatory
and the National Research Foundation (Grant No. 75415). RM is also
supported by the UK Science \& Technology Facilities Council (Grant
ST/S000550/1). This work made use of the South African Centre for High Performance Computing, under the project {\em Cosmology with Radio Telescopes,} ASTRO-0945.

\appendix

\section{Derivation of \eqref{eq:Deri_Dot}}\label{app1}
The scalar product of screen-space derivatives of two spherical harmonics, which is needed to obtain the lensing contribution to the bispectrum in \eqref{eq:Deri_Dot},  can be written  in terms of lowering and raising operators as \cite{DiDio:2015bua} 
\begin{equation}
\nabla_{\!\perp}^{a}Y_{\ell_3m_3}(\bm{n})\,\nabla_{\!\perp a}Y_{\ell_4m_4}(\bm{n})  = \frac{1}{2}\Big[\cancel\partial^{*} Y_{\ell_3m_3}(\bm{n})\,\cancel\partial Y_{\ell_4m_4}(\bm{n}) + \cancel\partial Y_{\ell_3m_3}(\bm{n})\,\cancel\partial^{*} Y_{\ell_4m_4}(\bm{n}) \Big].
\end{equation}
The effect of raising and lower operators on spherical harmonics is
\bea
\cancel\partial Y_{\ell m}&=&\sqrt{\ell(\ell+1)}~ _{1}Y_{\ell m}\label{eq:raise_eq}\\
\cancel\partial^{*} Y_{\ell m}&=&-\sqrt{\ell(\ell+1)}~ _{-1}Y_{\ell m}\,,
\label{eq:lower_eq}
\eea
where ${}_sY_{\ell m}$ are spin-weighted spherical harmonics, which obey  the product rule 
\begin{equation}
_{s_1}Y_{\ell_1m_1} {}_{s_2}Y_{\ell_2m_2}=\sum_{s\ell m}{} _sY_{\ell m}^{*}\sqrt{\frac{(2\ell_{1}+1)(2\ell_{2}+1)(2\ell+1)}{4\pi}}\begin{pmatrix}\ell_{1} & \ell_{2} & \ell\\
m_{1} & m_{2} & m
\end{pmatrix}\begin{pmatrix}\ell_{1} & \ell_{2} & \ell\\
-s_{1} & -s_{2} & -s
\end{pmatrix}.
\label{sysy}
\end{equation}
Using \eqref{eq:raise_eq} -- \eqref{sysy}, and symmetry properties of the Wigner 3j symbol, we obtain  \eqref{eq:Deri_Dot}.

\section{Lensing correction to the tree-level  4-point correlation function \label{sec:trispec}}

 We denote by $\mathcal{O}(n)$
a perturbation of order $n$. Assuming that $\mathcal{O}(1)$ perturbations
are Gaussian, the HI intensity mapping 4-point correlation function at tree level is $\big\langle\mathcal{O}(6)\big\rangle$,
i.e., $\big\langle\mathcal{O}(1)\mathcal{O}(1)\mathcal{O}(2)\mathcal{O}(2)\big\rangle$
and {$\big\langle\mathcal{O}(1)\mathcal{O}(1)\mathcal{O}(1)\mathcal{O}(3)\big\rangle$}.
We need to expand the
lensed HI fluctuations to fourth order: \begin{align}
\Delta^{L}(z,\bm{n}) & =\Delta\big(z,\bm{n}+\bm\nabla_{\!\perp}\phi(z,\bm{n})\big)\nonumber\\ 
 & =\Delta(z,\bm{n})+\sum_{m=1}^{4}\frac{1}{m!}\Big[\nabla_{\!\perp}^{a_{1}}\phi\cdots\nabla_{\!\perp}^{a_{m}}\phi\nabla_{\!\perp a_{1}}\cdots\nabla_{\!\perp a_{m}}\,\Delta\Big](z,\bm{n})-\mbox{average}\nonumber\\ 
 & =\Delta(z,\bm{n})+\sum_{m=1}^{4}L^{(m)}(z,\bm{n})-\big\langle L^{(2)}\big\rangle(z)-\big\langle L^{(4)}\big\rangle(z),\label{eq:lens_HI_expan}
\end{align}
where all orders of $\phi$ and $\Delta$ that add  to 4 or less
are included in the sum and we assume that the average of $\Delta$
has been removed. 

The lensing contribution to the  4-point correlation function only requires $L^{(2)}$ and $L^{(3)}$, and for these we need
the lensing potential up to second order. At first order the lensing potential is given by \eqref{lp}.
At second order, 
\begin{align}
\phi^{(2)}(z,\bm{n}) & =-2\int_{0}^{r}dr_{1}\frac{r-r_{1}}{rr_{1}}\varphi^{(2)}(z_{1},\bm{n})-2\int_{0}^{r}dr_{1}\frac{r-r_{1}}{rr_{1}}\nabla_{\!\perp}^{a}\phi^{(1)}(z_{1},\bm{n})\nabla_{\!\perp a}\varphi^{(1)}(z_{1},\bm{n})\,,
\end{align}
where $\varphi=(\Phi+\Psi)/2$ and
$r_{i}\equiv r(z_{i})$. Here the terms which include $\nabla_{\!\perp}^{a}\varphi$
are the so-called post-Born terms, since they take into account the
fact that the photon is not propagating along the unperturbed direction
$\bm{n}$. 
With these expressions, the lensing correction to $\Delta$ at third  order
 is
\begin{align}
L^{(3)} & =\nabla_{\perp}^{a}\phi^{(1)}\,\nabla_{\perp a}\Delta^{(2)}+\frac{1}{2}\nabla_{\perp}^{a}\phi^{(1)}\nabla_{\perp}^{b}\phi^{(1)}\,\nabla_{\perp a}\nabla_{\perp b}\Delta^{(1)}+\nabla_{\perp}^{a}\phi^{(2)}\nabla_{\perp a}\Delta^{(1)}\,,\label{eq:Lens3_Order}
\end{align}
In a $\Lambda$CDM cosmology at late times, the Weyl potential and
the metric potentials are equal at first order: $\varphi^{(1)}=\Phi^{(1)}=\Psi^{(1)}$.
This also holds at higher order on sub-Hubble scales. Furthermore,
the screen-space Laplacian of $\varphi$ is well approximated by the
3D Laplacian on sub-Hubble scales and the Poisson equation maintains
its Newtonian form. This implies that 
\begin{equation}
\nabla_{\!\perp}^{2}\varphi^{(n)}\simeq\nabla^{2}\varphi^{(n)}\simeq\nabla^{2}\Phi^{(n)}\simeq\frac{3}{2}\Omega_{m}\mathcal{H}^{2}\delta^{(n)}
\end{equation}
where $\Omega_{m}\mathcal{H}^{2}=\Omega_{m0}H_{0}^{2}/a$ and $\delta^{(n)}$
is the Newtonian density contrast \cite{Bernardeau:2001qr}. The lensed
4-point correlation function  in redshift space is written as
\begin{equation}
T^{\rm L}(z_i,\bm n_i) =\big\langle\Delta_{1}^{\rm L}\Delta_{2}^{\rm L}\Delta_{3}^{\rm L}\Delta_{4}^{\rm L}\big\rangle+\delta T(z_i,\bm n_i).
\end{equation}
Here $\delta T$ is the lensing correction to the unlensed $T$. At tree level, the 4-point correlation function is of the form $\langle O(1)O(1)O(1)O(3)\rangle+\langle O(1)O(1)O(2)O(2)\rangle$.
In detail
\begin{align}
T^{\rm L}(z_i,\bm n_i) & =\big\langle\Delta_{1}^{(1)}\Delta_{2}^{(1)}\Delta_{3}^{(1)}\big[\Delta_{4}^{(3)}+L_{4}^{(3)}\big]\big\rangle+\text{3 perms}\nonumber \\
 & +\big\langle\Delta_{1}^{(1)}\Delta_{2}^{(1)}\big[\Delta_{3}^{(2)}+L_{3}^{(2)}-\big\langle L_{3}^{(2)}\big\rangle\big]\big[\Delta_{4}^{(2)}+L_{4}^{(2)}-\big\langle L_{4}^{(2)}\big\rangle\big]\big\rangle+\text{5 perms}.
\end{align}
The tree-level lensing correction is thus made up of two parts:
\begin{align}
\delta T[1](z_i,\bm n_i) & =\big\langle\Delta_{1}^{(1)}\Delta_{2}^{(1)}\Delta_{3}^{(1)}L_{4}^{(3)}\big\rangle+\text{3 perms}, \label{eq:trispec_first}\\
\delta T[2](z_i,\bm n_i) & =\big\langle\Delta_{1}^{(1)}\Delta_{2}^{(1)}\Delta_{3}^{(2)}L_{4}^{(2)}\big\rangle-\big\langle\Delta_{1}^{(1)}\Delta_{2}^{(1)}\Delta_{3}^{(2)}\big\rangle\big\langle L_{4}^{(2)}\big\rangle+\big\langle\Delta_{1}^{(1)}\Delta_{2}^{(1)}\big\rangle\big\langle L_{3}^{(2)}\big\rangle\big\langle L_{4}^{(2)}\big\rangle\nonumber \\
 & \ +\big\langle\Delta_{1}^{(1)}\Delta_{2}^{(1)}L_{3}^{(2)}\Delta_{4}^{(2)}\big\rangle+\big\langle\Delta_{1}^{(1)}\Delta_{2}^{(1)}L_{3}^{(2)}L_{4}^{(2)}\big\rangle-\big\langle\Delta_{1}^{(1)}\Delta_{2}^{(1)}L_{3}^{(2)}\big\rangle\big\langle L_{4}^{(2)}\big\rangle\nonumber \\
 & \ -\big\langle\Delta_{1}^{(1)}\Delta_{2}^{(1)}\Delta_{4}^{(2)}\big\rangle\big\langle L_{3}^{(2)}\big\rangle-\big\langle\Delta_{1}^{(1)}\Delta_{2}^{(1)}L_{4}^{(2)}\big\rangle\big\langle L_{3}^{(2)}\big\rangle+\text{5 perms}.
\end{align}

First consider
$\delta T[1]$, which is obtained by using \eqref{eq:Lens3_Order} in \eqref{eq:trispec_first}:
\begin{align}
\delta T[1](z_i,\bm n_i) & =\big\langle \Delta_{1}^{(1)}\Delta_{2}^{(1)}\Delta_{3}^{(1)}\nabla_{\perp}^{a}\phi_{4}^{(1)}\Delta_{\perp a}\Delta_{4}^{(2)}\big\rangle+\frac{1}{2}\big\langle\Delta_{1}^{(1)}\Delta_{2}^{(1)}\Delta_{3}^{(1)}\nabla_{\perp}^{a}\phi_{4}^{(1)}\nabla_{\perp}^{b}\phi_{4}^{(1)}\nabla_{\perp a}\nabla_{\perp b}\Delta_{4}^{(1)}\big\rangle\nonumber \\
 & \ +\big\langle \Delta_{1}^{(1)}\Delta_{2}^{(1)}\Delta_{3}^{(1)}\nabla_{\perp}^{a}\phi_{4}^{(2)}\Delta_{\perp a}\Delta_{4}^{(1)}\big\rangle \ +\text{3 perms}.
\end{align}
The other terms can  be similarly obtained.

\clearpage
\bibliographystyle{JHEP}
\bibliography{bibliography}
 
\end{document}